# Spontaneous Liquid Crystal and Ferromagnetic Ordering of Colloidal Magnetic Nanoplates

M. Shuai,[1] A. Klittnick,[1] Y. Shen,[1] G. P. Smith,[1] M. R. Tuchband,[1] C. Zhu,[2] R. G. Petschek,[3] A. Mertelj,[4] D. Lisjak,[4] M. Čopič,[4,5] J. E. Maclennan,[1] M. A. Glaser,[1] N. A. Clark[1,*]

**Affiliations:**

[1]Department of Physics and Soft Materials Research Center, University of Colorado, Boulder, Colorado 80309, USA.

[2]Advanced Light Source, Lawrence Berkeley National Laboratory, Berkeley, California 94720, USA.

[3]Physics Department, Case Western Reserve University, Cleveland, Ohio 44106, USA.

[4]Jozef Stefan Institute, SI-1000 Ljubljana, Slovenia.

[5]Faculty of Mathematics and Physics, University of Ljubljana, SI-1000 Ljubljana, Slovenia.

*Correspondence to: Noel.Clark@colorado.edu

**Abstract**: Ferrofluids are familiar as colloidal suspensions of ferromagnetic nanoparticles in aqueous or organic solvents. The dispersed particles are randomly oriented but their moments become aligned if a magnetic field is applied, producing a variety of exotic and useful magneto-mechanical effects. A longstanding interest and challenge has been to make such suspensions macroscopically ferromagnetic, that is having uniform magnetic alignment in the absence of a field. Here we report a fluid suspension of magnetic nanoplates which spontaneously aligns into an equilibrium nematic liquid crystal phase that is also macroscopically ferromagnetic. Its zero-field magnetization produces distinctive magnetic self-interaction effects, including liquid crystal textures of fluid block domains arranged in closed flux loops, and makes this phase highly sensitive, with it dramatically changing shape even in the Earth's magnetic field.

**Introduction**

Colloidal suspensions of nanoparticles offer exciting and ever-expanding opportunities for creation and application of novel hybrid particle/fluid phenomena. A classic example is the paramagnetic ferrofluid, a colloidal suspension of ferromagnetic nanoparticles dispersed in isotropic solvent[1,2]. The ferrofluid particles disperse with random orientations but readily align and move in magnetic fields, resulting in dazzling hybrid behaviors, including radical changes in shape[3,4] and magnetically controlled transport, which enable a variety of technical and biomedical applications[5,6]. A longstanding challenge has been to make fluids that are ferromagnetic, such that they exhibit spontaneous, equilibrium magnetic ordering in the absence of an applied field[7-11]. Such phases are conceivable because in such hybrid suspensions the dipole-dipole interaction energy between the supermolecular magnetic moments of neighboring particles can be many times the thermal energy at room temperature[10,11].

One approach to achieve ferromagnetism in colloidal fluid is to create stable, fluid suspensions of well dispersed magnetic nanoparticles in isotropic solvent, and design their mutual interactions to produce equilibrium, zero-field magnetization. Since ferromagnetic ordering implies orientational ordering of the particles, such fluid suspensions are also liquid crystals (LCs). Suspensions of magnetic nanoplates with moments normal to their planes are of particular interest to this study. One reason is that anisotropic steric interactions can lead to spontaneous LC ordering[12], which have been demonstrated in various colloidal suspensions of nanoplates and nanosheets[13,14]. Meanwhile, the plate-like (as opposed to spherical or rod-like) shape promotes parallel rather than antiparallel orientation of the moments of neighboring particles, making the interplate magnetic dipole interactions sufficient to stabilize ferromagnetic ordering, as recently demonstrated in dilute suspensions of nanoplates orientationally ordered by a thermotropic nematic LC host fluid[15,16].

In our experiments, barium hexaferrite (BF) nanoplates are suspended in isotropic *n*-butanol and surfactant-stabilized to produce a system of functionalized nanoplates with weak electrostatic repulsion, strong and anisotropic steric repulsion, and magnetic interaction. The electrostatic repulsion enables stable suspensions at essentially any concentration, including at high volume fractions where spontaneous LC ordering may occur. We observe nematic ordering of the nanoplates at volume fractions in the range where the Onsager model and subsequent theory and simulations predict an isotropic (Iso) – nematic transition for polydisperse, plate-like colloids[17-19]. This nematic is distinctly ferromagnetic, with a magnetization density two orders of magnitude larger than that reported in the previously found thermotropic nematic LC host (4-cyano-4'-pentylbiphenyl, 5CB) fluid system[15].

**Results**

*Isotropic to ferromagnetic nematic phase transition*

Suspensions of BF nanoplates (overall thickness $t = 7$ nm, diameter $D = 48\pm21$ nm, Supplementary Fig. 1) in *n*-butanol (BF/BuOH) are loaded in capillaries and observed at various magnifications using both unpolarized and depolarized transmitted light microscopy, which enabled visualization of their basic phase behavior as a function of the volume fraction of the functionalized nanoplates, $\phi = \pi \langle D^2 t \rangle \nu/4$, where $\nu$ is the number density (Fig. 1a). For $\phi < 0.28$, the suspensions are optically isotropic, giving excellent extinction for any sample orientation between crossed polarizer and analyzer, and uniform transmission with only the polarizer, indicating that the particles are well dispersed and that the suspensions are optically homogeneous and isotropic on micron and larger scales. The characteristic orange-red color is due to nanoplate absorption (Supplementary Fig. 2). The isotropic suspensions are stable, with only weak gravitational concentration gradients developing over periods of hundreds of days (Supplementary Fig. 3), consistent with an estimated Perrin length $l_P = k_B T/\Delta\rho_{LS}(\pi \langle D^2 t \rangle/4)g \sim 1$ cm ($\Delta\rho_{LS}$ = mass density difference between nanoplates and solvent, $g = 9.8$ m s$^{-2}$).

With application of an external magnetic field $\mathbf{B}_{ext}$, the Iso suspensions exhibit induced birefringence, with significant optical transmission between crossed polarizers when $B_{ext} \sim 10$ mT is applied in the plane of the cell at angle of 45 º to the polarizer, **P** (Fig. 1b). This induced birefringence is spatially homogeneous, further evidence that the nanoplate dispersion is uniform on the optical scale. The direction with lowest refractive index is found using a Berek optical compensator to be parallel to $\mathbf{B}_{ext}$. The Iso suspensions also exhibit magnetically-induced optical dichroism, with the absorbance largest for optical polarization parallel to the planes of the nanoplates (Supplementary Fig. 2). Thus, the magnetized isotropic phase has field-induced, uniaxial nematic ordering of the plates, with the nematic director $\mathbf{n}(\mathbf{r})$, the unit vector along the local, mean orientation direction of the nanoplate normals, aligned parallel to $\mathbf{B}_{ext}$, along the induced magnetization density **M**. The induced nematic has negative birefringence $\Delta n(B_{ext}) = n_\parallel -$ $n_\perp < 0$ and a visible light dichroic ratio $\zeta(B_{ext}) = OD_\parallel/OD_\perp < 1$ [where $\parallel$ ($\perp$) indicates orientation parallel (normal) to **n**].

At higher concentrations ($\phi \gtrsim 0.28$), fresh suspensions begin to sediment higher-density domains at the bottom of the capillary that are birefringent even in the absence of applied magnetic field. Over periods of hours to days, a well-defined, horizontal, planar interface between the birefringent region and the upper isotropic phase develops. The birefringent region, if left unperturbed, gradually develops LC-like textures with uniform domains that appear to grow in size and become more ordered with the passage of time (Fig. 1a,c, below the interface, and Supplementary Fig. 4). We identify this behavior as two-phase coexistence, with the denser phase sedimented from the less-dense Iso phase by gravity.

Further insight into the ordered phase can be obtained by observing the Iso/$N_F$ interface with an external field applied, where we see that an initially sharp interface formed in the absence of field (Fig. 1c) begins to deform and move into the Iso region as **B**$_{ext}$ is increased. For sufficiently large field ($B_{ext} \sim 0.5$ mT), the ordered phase reorients to have the director along **B**$_{ext}$, nematic order induced in the Iso phase becomes comparable to that of the ordered phase, and the interface becomes continuous (Fig. 1d), qualitatively as described by the mean-field model of a field-induced isotropic/nematic transition of Fan and Stephen[20]. Induction of nematic order requires larger **B**$_{ext}$ as the interface is pushed further up in the cell because the number density of nanoplates in the Iso region decreases with increasing height (Supplementary Fig. 3). If the applied field is removed, the interface reforms and returns to the state of Fig. 1c. These responses to changing field take several minutes to hours to complete because the distribution in the number density of plates must also change to reestablish a uniform chemical potential of the plates in the colloidal system. The field-induced continuity provides evidence that the birefringent phase below the interface is identical to that induced in the Iso phase, a uniaxial, magnetically polar nematic, having the first- and second-rank nanoplate orientational order parameters $Q_1$ and $Q_2$ simultaneously non-zero (ferromagnetic nematic, $N_F$). If **B**$_{ext}$ is instead reduced to a small but non-zero value, the Iso/$N_F$ interface was also found, unexpectedly, to exhibit a field-induced spiking instability, discussed in detail below.

In order to assess the nature of the magnetic ordering in the vicinity of the Iso/$N_F$ transition, we measured and analyzed $\Delta n$ induced by an external magnetic field $B_{ext}$, in effect probing the field-induced, uniaxial orientational order in the Iso and $N_F$ phases (Supplementary Fig. 5). Analysis including the effects of polydispersity of the response of extremely low-concentration suspensions ($\phi = 0.005$) gives an estimate of the mean nanoplate magnetization of $m_o = 2 \times 10^{-18}$ A m$^2$ (see Supplementary Note 1), which is very close to the value obtained from aligned and dried BF nanoplates with a mean diameter of 70 nm[15]. As $\phi$ is increased, the initial (low-field) magnetic susceptibility, $\chi(\phi)$, of the Iso phase, derived from the measurements of $Q_2 = \Delta n/\Delta n_{sat}$, where $\Delta n_{sat}$ is the saturated birefringence, shows a substantial growth. Comparison shows that the standard Langevin-Weiss mean-field model[21] significantly overestimates the growth of $\chi(\phi)$ and predicts the Iso/$N_F$ transition near $\phi = 0.09$, a much lower value than observed. However, treating the orientationally diffusing nanoplates of the Iso phase effectively as a system of dipolar hard spheres and applying models that include interparticle correlations[11] gives a reasonable qualitative description of the Iso phase susceptibility (see Supplementary Note 2).

The LC order of the $N_F$ phase was characterized by synchrotron X-ray diffraction. A high-concentration sample ($\phi = 0.28$) was kept vertical for several days until it developed a clear Iso/ $N_F$ interface. As shown in Fig. 2a, with the beam intersecting the upper part of the Iso part of the sample (*Location Iso*), the diffraction pattern is circularly symmetric, demonstrating the isotropic nature of the sample. When the beam is incident on the two-phase-interface (*Location Iso/$N_F$*), a pair of diffuse arcs appears, with a noticeable isotropic background. The higher-density phase region (*Location $N_F$*) gives multiple diffraction arcs. If the diffracted signal is integrated along concentric circles around the beam, we obtain the dependence of scattering intensity versus wavevector $q$ shown in Fig. 2b. The diffraction arcs are broad, indicating short-ranged lamellar correlations characteristic of a nematic phase[22], with the peaks at almost the same locations in both regions of the sample. Fitting the structure factor $S(q) = q^2I(q)$ with the function $S(q) = \text{sh}(q\alpha)/[\text{ch}(q\alpha)-\cos(qd)]$ ($\alpha$ is the average of the magnitude of the orientation difference between neighboring plates), describing lamellar correlations in an array of plates, gives peaks in the pair correlation function at $d = 20.0$ nm, 20.9 nm, and 22.1 nm for the $N_F$, Iso/$N_F$, and Iso regions respectively (Fig. 2c), showing that the difference in the number density between the two phases is small. The azimuthal distributions of the first-order diffraction intensities at *Location Iso/$N_F$* and *Location $N_F$* are plotted in Fig. 2d. Analysis of the azimuthal distributions of the diffraction intensity gives a second-rank order parameter $Q_2 = 0.8$ (see Supplementary Note 3).

*Equilibrium magnetic domain structure in a thin capillary*

The response of the higher-density phase to applied magnetic field was studied in some detail. This phase is a low-viscosity fluid that flows readily if the capillary is tilted and anneals if left undisturbed into domains that are several millimeters across and span the 50 μm thickness of the capillary (Fig. 3 and Supplementary Fig. 4). Cells were equilibrated over periods of up to several days with external static fields $\mathbf{B}_{equil}$. If $B_{equil} = 0$, obtained by locally cancelling the Earth's magnetic field and other stray fields, a characteristic texture of large, distinctly dichroic and birefringent domains which divide the cell into almost uniform regions of different orientation separated by sharp domain boundaries as shown in Fig. 3, consistently develops. Measurement of the dichroism and birefringence of these domains and applying the relationship between **n**, $\Delta n$, and $\zeta$ discussed above in connection with the Iso phase, shows that the nanoplate director **n**(**r**) is oriented as indicated in Fig. 3a. If $\mathbf{B}_{equil}$ is applied parallel to **n** in the left- and right-hand domains in Fig. 3a, then one of these domains (in this case the left domain) shrinks and the other expands (Fig. 3b). If the sign of $\mathbf{B}_{equil}$ is reversed then the domain growth/shrinkage is also reversed. The field-induced displacement of the inversion walls is plotted in Supplementary Fig. 6. These walls behave as a system of lines with inherent tension that adopt a geometry minimizing their net length while mediating the changes in director orientation. The field-induced wall displacements are a consequence of stresses that tend to change the domain areas according to the sign of the magnetic energy density $U_M = -\mathbf{M} \cdot \mathbf{B}_{equil}$. The linear response of the displacement at low fields is definitive evidence for a magnetization, **M**, that is finite and independent of applied field strength around $B_{equil} = 0$. The field-induced motion of the domains walls is maximal for $\mathbf{B}_{equil}$ in the **n** direction, showing that **M** is along **n** and uniquely determining its sign, as indicated in Fig. 3a. Equilibration in a finite field, such as the Earth's magnetic field, $B_{equil} \sim 0.05$ mT, produces arrays of similarly structured domains that are more complex (Supplementary Fig. 4).

The domain magnetization is also evident from the internal response of the domains to applied fields $\mathbf{B}_{trans}$ changing over periods of several seconds. The big domains in the cell are observed to have a static, grainy texture of subdomains ~10 μm in size in which the in-plane orientation varies around the mean by a few degrees. Applying $\mathbf{B}_{trans}$ parallel to $\mathbf{M}$ within a domain produces a continuous enhancement of the extinction of that domain when viewed between crossed polarizers as the magnitude of the applied field is increased (Fig. 3c). The field applies torques to the nanoplates that tend to reduce the orientational fluctuations of the director field $\mathbf{n}(\mathbf{r})$, orienting the plates more parallel on average to $\mathbf{B}_{trans}$ and thereby increasing $Q_2$. Applying $\mathbf{B}_{trans}$ antiparallel to $\mathbf{M}$, on the other hand, generates torques that tend to enhance the orientational disorder of the nanoplates, producing, above a certain threshold $B_{trans} \sim 0.005$ mT, an instability marked by the appearance of a speckle-like pattern of birefringence generated by an array of subdomains of $\mathbf{M}(\mathbf{r})$ that have rotated away from the field direction in random directions and through a distribution of angles. Since these subdomains are substantially smaller than the cell thickness, they must be a response to static fluctuations and thus to quenched disorder in the local $\mathbf{M},\mathbf{n}$ fields. Strong magnetic self-interactions lead to the local expulsion of magnetic charge and the creation of subdomains responding as independent units to $B_{ext}$. Since these changes take place at values of $B_{ext}$ much smaller than the internal field ($B_M \sim 40$ mT, see Supplementary Note 4), the response appears to be a soft deformation mode in the frozen transverse fluctuations of $\mathbf{M}$. This behavior is quite different from the Fréedericksz transitions found in the BF/5CB system[23].

This dependence of the domain response on the sign of $\mathbf{B}_{trans}$, combined with imaging of the dichroism and birefringence-based optical textures, enables detailed mapping of $\mathbf{M}(\mathbf{r})$ and $\mathbf{n}(\mathbf{r})$ in any domain pattern. $\mathbf{M}(\mathbf{r})$ is found to be always parallel to $\mathbf{n}(\mathbf{r})$, which is consistent with the nanoplate magnetic moments being along their normals, and with the observed domain wall movements in applied fields. Importantly, the domain magnetization is non-zero in the absence of any applied magnetic field, as an equilibrium property of the phase.

The textural features and response to applied field of the BF/BuOH ferromagnetic nematic phase may be understood as phenomena resulting from the combined effects of magnetostatic and nematic elastic energies. The interplay of Frank elasticity and magnetostatic self-interactions is expressed through the characteristic length $\xi_M = \sqrt{K/(\mu_o M^2)} < 1$ μm ($K$ is the LC Frank elastic constant in the one-constant approximation, see Supplementary Fig. 7 and Supplementary Note 4), which for the BF/BuOH $N_F$ phase is small compared to any cell dimension. Generally, the imperative produced by this condition is to reduce magnetostatic energy by reducing the excess internal and external magnetic fields generated. A distortion such as splay of **M**(**r**) that produces bulk magnetic charge of particular sign [$\rho_m(\mathbf{r}) = -\nabla \cdot \mathbf{M}(\mathbf{r})$], must necessarily also produce the same quantity of opposite charge elsewhere in the bulk or at the interfaces with non-magnetized material [$\rho_s(\mathbf{r}) = \mathbf{M}(\mathbf{r}) \cdot \mathbf{s}(\mathbf{r})$]. Therefore, the drive for energy reduction leads either to non-splayed regions of **M**(**r**) tangent to boundaries, which produce minimal magnetic charge, or to orientational variation of **M**(**r**) such that the positive and negative magnetic charges generated are located as close as possible to each other, a separation that is at least $\xi_M$ because of the LC elasticity (Supplementary Fig. 7). The resulting structure of uniform domains separated by distinct walls is a general feature of the BF/BuOH textures reported here. However, in systems wherer $\xi_M$ is much larger than the cell dimensions, such as in the case of BF/5CB, the magnetic self-interactions affect the texture only weakly (see Supplementary Note 5).

Our observations indicate that the $N_F$ region in the flat capillary cells is magnetized with **M** preferentially parallel to the largest cell face (Fig. 3d). This is expected considering that for a slab of infinite area, the magnetic energy of a uniform **M** field free to orient in any direction is lowest when **M** is in the plane of the slab (see Supplementary Note 4). The response of the capillary cells to applied field confirms this confinement effect (Supplementary Fig. 8), showing that, while fields of less than 0.2 mT were needed to induce a complete in-plane alignment of the entire sample, in the case of perpendicular field there is little observable response until $B_{ext\perp}$ reaches about 10 mT, and extinguishing regions first appear at about 20 mT, indicating a domain-mediated transition orienting **n** normal to the cell face.

The equilibrium in-plane texture in the absence of external applied fields can be understood by approximating the $N_F$ region as a two-dimensional (2D) slab the thickness of the cell, in which **n**(**r**) is a 2D field, uniform along the cell normal, **c** (Fig. 3d), and assuming that boundary conditions force **n**(**r**) to be parallel to the edges of the cell, which eliminates magnetic charge on the surfaces. **M**(**r**), thus constrained to be parallel to the **a**,**b** plane in Fig. 3d, adopts an equilibrium in-plane texture that is determined only by the in-plane sample shape and the magneto-elastic behavior. In the case of rectangular cells with planar boundaries, the equilibrium configuration is a loop of uniform domains of **M** separated by sharp splay/bend/splay walls, such as that shown in Fig. 3e and Supplementary Fig. 7, forced into the texture of Fig. 3a by the bends of **M(r)** at the corners of the rectangle. These walls are observed to have an optical-resolution (micron scale or smaller) linear core and a diffuse width comparable to the cell thickness. Figure 3e shows that the deformation of **M(r)** through a splay/bend/splay wall generates lines of opposite magnetic charge, the attraction of which favors narrow walls. The Frank elastic energy associated with bend of the director **n(r)**, on the other hand, favors broad walls, in order to reduce the curvature of the director reorientation at the wall center[24]. Using $K_S \sim 6k_BT/D = 5\times10^{-13}$ N from a numerical simulation[25], this balance yields an equilibrium wall core structure of width $w \sim \xi_M < 1$ μm, which is smaller than the optical resolution. This scenario is an example of "orientational fracture", similar to that found in -1 topological defects in the director field structure and textures of thermotropic ferroelectric smectic C LCs with large permanent polarization[26], as detailed in Supplementary Note 6.

The loop geometry leads to an overall topological disclination line defect of strength +1 in the **M,n** fields of the sample. In the center of the cell we observe a linear boundary between domains of antiparallel **M** (Fig. 3a-c), which higher magnification imaging shows to be a +1 disclination line singularity in the **M,n** fields (Fig. 3f) that starts on one cell surface at a point where three domains meet (blue square) and ends on the other cell surface at the other such point, as shown in Fig. 3d. This defect can be constructed by starting with a +1 bend disclination line (blue box) running along **c** from one cell face to the other (required by the boundary conditions on the cell edges), and then sliding the intersection points on the cell surfaces to the locations where the three domains meet (blue squares). This defect will "escape" into the third dimension, forming a continuous twist wall in the cell center (orange square) confined by the mutual attraction of the opposite-signed magnetic space charge.

The equilibrium, field-free domain pattern of **M(r)** in Fig. 3 resembles that of the closed flux loops found in solid ferromagnetic single crystals[27], known to be an organization of their magnetization field **M(r)** that reduces magnetic charge $\rho_m(\mathbf{r}) = -\nabla \cdot \mathbf{M}(\mathbf{r})$ and thereby the magnetic field and energy external to the sample. In crystals, such patterns are achieved with the aid of the crystal anisotropy, which pins **M(r)** to certain "easy" symmetry directions. In ferromagnetic liquid crystals, however, there is no underlying lattice to stabilize discontinuous reorientation of **M(r)**. Equilibrium domain patterns are established by the sample shape and by magneto-elastic interactions described in Supplementary Note 4. An interesting special case where the magnetization can vary continuously with no need for domain walls is the ring cell, for example one shaped like a flat washer or a racetrack. The equilibrium **M(r)** here is everywhere parallel to the cell boundary, a geometry with continuous bend depending only on radius.

*Spikes at the Iso/$N_F$ interface*

A dramatic manifestation of the fluid ferromagnetic ordering in the BF/BuOH system is the interfacial spiking behavior shown in Fig.4 and Supplementary Fig. 4. In fresh cells, the Iso/$N_F$ interface is initially flattened by gravitational sedimentation of domains of $N_F$ (Fig. 1c). However, applying an external magnetic field of $B_{equil} \gtrsim 0.005$ mT normal to this interface, causes a quasi-periodic array of birefringent spikes to grow which equilibrate to a field-dependent height above the interface over the course of a day (Fig. 4a,b). Exposure of the spikes to transient fields of few-second duration shows that they are nematic and ferromagnetic, with magnetization parallel to the spike axis along **B**$_{equil}$, with a field response very similar to that of the bulk $N_F$ domain.

The spike geometry appears to be a soliton-like response of the interface that enables the magnetization of a ferromagnetic fluid to align with **B**$_{ext}$. The extended shape of the spikes, with **M(r)** oriented along the spike axis, minimizes the generation of magnetic charge, with the remnant opposite magnetic charges on opposite ends of the spikes, indicated in Fig. 4c, decreasing as the spikes become longer and narrower. Each spike is effectively a small bar magnet, making the spikes mutually repulsive, as is the case for similarly aligned bar magnets placed next to one another. The spikes are roughly conical in shape, becoming narrower in both the **a** and **c** directions with increasing distance from the interface, the latter case evident from the birefringent interference extinctions at places where the phase shift is an odd multiple of $\pi$.

Spikes tend to grow in an external field if the magnetic energy gain is larger than the cost of increasing the interface area, *e.g.*, for a cylindrical spike of radius $R \sim 25$ μm when $B_{ext} > 2\sigma_{IN}/RM \sim 0.02$ mT. As discussed in connection with the Iso phase, application of a sufficiently large $B_{ext}$ normal to the interface of an equilibrated sample induces nematic order everywhere, expanding the strong ordering of the bulk nematic well into the Iso region for $B_{ext} \gtrsim 0.5$ mT (Fig. 1d). Upon the removal of $B_{ext}$ from such a sample previously equilibrated with spikes, both the interface and the spikes evolve quickly, over a period of a few minutes, as shown in Fig. 4d, to an instability of interface position. This experiment shows that the spike length at a given field is limited by the decrease in the susceptibility to magnetic ordering with distance from the Iso/$N_F$ interface.

We propose that spike formation can be understood qualitatively as a novel type of nonlinear capillary instability of the sort produced in fluid interfaces when they become electrically or magnetically charged by the application of an external field[3,5,28]. As discussed above, in a ferromagnetic fluid, the permanent magnetization **M** is parallel to the interface when $B_{ext} = 0$, but as **M** rotates in response to an applied field normal to the interface, it develops a normal component $\mathbf{M}_\perp = \mathbf{B}_{ext}/\mu_o$. This, in turn, produces magnetic charge $B_{ext}/\mu_o$, and an effective magnetization energy density for driving the instability. Near the threshold, the instabilities are periodic with wavelength $\lambda_G = 2\pi\, l_G$, where $l_G = \sqrt{\sigma/(\Delta\rho g)}$ is the gravitational capillary length ($\sigma$ = surface tension, $\Delta\rho$ = mass density difference between Iso and $N_F$ phases). The associated increase in capillary energy density[3] is $U_G = \sqrt{\sigma\Delta\rho g}$. Taking $\sigma_{IN} = 0.1 k_B T/D^2 \sim 10^{-7}$ J m$^{-2}$ as an estimate of the Iso/$N_F$ interfacial tension[29,30], we find $\lambda_G \sim 200$ μm, comparable to the observed spike spacing, and $U_G \sim 0.002$ J m$^{-3}$, energy that is provided by the applied field.

**Discussion**

A variety of observations discussed above indicate that the Iso and $N_F$ phases are equilibrium dispersions of individual nanoplates. First, the magnitude of the magnetization of the orientable unit producing the magnetic birefringence in the Iso phase is near that of a single plate, which must therefore act as thermalized individuals. Second, the Iso/$N_F$ phase coexistence appears at $\phi \sim 0.28$, equivalent to a scaled density of $\nu\langle D^3\rangle = 3.3$, a volume fraction value where nematic ordering due to the steric and charge interactions of single nanoplates is expected[18,19,31]. Third, XRD from oriented monodomains of the $N_F$ phase reveals several diffuse quasi-Bragg reflections from short-ranged lamellar ordering with a layer spacing of $d = 20$ nm (Fig. 2). Assuming that the nanoplates form nearly continuous sheets, *i.e.*, the volume fraction $\phi = 0.28$ is determined by the spacing between the neighboring plates, it gives a plate thickness $t = \phi d = 5.6$ nm. This is comparable to the 7 nm thickness of a single nanoplate, providing further evidence that the BF/BuOH system is a suspension of single plates dispersed in equilibrium and distinguishing the ferromagnetism observed here from that of the non-ergodic aggregation of magnetic particles[8,9]. Finally, observations above suggest that magnetic dipolar and steric quadrupolar interactions, tending to stabilize $Q_1$ and $Q_2$, respectively, both contribute to the ordering transition.

Several theoretical treatments of ordering transitions in systems with both dipolar and quadrupolar interactions have explored the interplay of the ordering of $Q_1$ and $Q_2$, in particular the Sivardiere and Blume (SB) Ising model in the molecular field approximation[32], and the Baus and Colot (BC) density functional theory of a system of dipolar hard ellipsoids[33]. Despite the very different nature of these two systems, the SB and BC phase diagrams, shown in Supplementary Fig. 9, exhibit some key qualitative common features relevant to the BF/BuOH system. Both the SB and BC show that the two types of ordering are surprisingly weakly coupled, with the isotropic/nematic quadrupolar transition independent of the dipole interaction until the latter becomes strong enough for dipole ordering to occur at a smaller density than the quadrupolar ordering. On the other hand, the dependence of the (isotropic or nematic)/$N_F$ transition on the dipole strength is only weakly affected by whether the quadrupolar ordering has taken place or not. Our independent assessment above of the proximity of the two transitions at $\phi \sim 0.28$ leads us to conclude that on these phase diagrams the BF/BuOH system is in the vicinity of the tricritical point where, with increasing dipole strength, the nonferromagnetic nematic phase has just disappeared. This is a place where the remnant Iso/$N_F$ transition will be first order, and where there will be significant short-ranged quadrupolar ordering (nonpolar alignment of the nanoplates) and thus an enhanced susceptibility for inducing $Q_2$.

Those familiar with smectic liquid crystals may recognize the similarity of the phase diagram in Supplementary Fig. 9 to McMillan mean-field isotropic/nematic/smectic A model, with ferromagnetic ordering replacing the appearance of smectic A layering[34]. McMillan's distinct mean-field approach is thus yet another route to the generic dipolar/quadrupolar phase diagram. The Landau model of Palffy-Muhoray, Lee, and Petschek also produces a similar Iso, nematic, $N_F$ phase diagram[35].

**Methods**

*Preparation of the BF suspensions*

Barium hexaferrite nanoplatelets were suspended in *n*-butanol to form a stable colloidal suspension following a previously established method[36,37]. Briefly, BF nanoplatelets with the nominal composition $BaFe_{11.5}Sc_{0.5}O_{19}$ were synthesized hydrothermally at 240°C with dodecylbenzenesulphonic acid (DBSa) added as a surfactant. The nanoplatelets were subsequently washed with water and nitric acid. To increase the adsorption of the DBSa onto the nanoplatelets, they were suspended in an $HNO_3$ solution (pH = 1.5) and heated at 100 °C for 2.5 h. After this the nanopalatelets were washed again with water and acetone, dried in air at 60 °C and dispersed in n-butanol by probe sonication. The final volume fraction of platelets $\phi = 0.003$ was obtained by concentrating the initial suspension ($\phi = 0.001$) in a rotary evaporator. The functionalized nanoplatelets have an average magnetic core thickness of 5 nm, with 1 nm of surfactant on each side, and a mean diameter of $D = 48$ nm with a standard deviation of 21 nm, which gives an aspect ratio of ~ 7. Suspensions prepared at the volume fraction obtained in the synthesis of the platelets, $\phi = 0.003$, formed an isotropic liquid. Higher concentrations were prepared by compression, using ultra-centrifugation at 160,000 g for 40 min. The clear supernatant was then removed and the remaining concentrated suspension, at $\phi \sim 0.1$, homogenized. Repeated addition of low-concentration suspension followed by recompression eventually yields suspensions with desired higher concentrations. The suspensions, loaded into the 50 μm thick ×1 mm wide ×25 mm long sample cavity of rectangular capillaries, flame-sealed, and held vertically, are stable for months under ambient conditions, showing no sedimentation or aggregates visible in the optical microscope.

*Synchrotron X-ray experiment*

The orientational order of the nanoplates in a phase-separated suspension was studied by synchrotron X-ray diffraction performed on Beamline 7.3.3 of the Advanced Light Source, Berkeley, CA, USA. The suspensions were contained in rectangular (0.05 ×1.0 $mm^2$) borosilicate glass capillaries. The sample was probed at multiple locations in the capillary with a 10 keV synchrotron X-ray beam with a Mo/$B_4$C double-multilayer monochromator and a beam at the sample about 700 μm wide by 300 μm high.

**Acknowledgements**

This work was supported by the Soft Materials Research Center under NSF MRSEC Grants DMR-0820579 and DMR-1420736, by Institute for Complex Adaptive Matter Postdoctoral Fellowship Award OCG5711B, and by Slovenian Research Agency Grants P1-0192 and P2-0089-4. The synchrotron X-ray experiments weresupported by the Director of the Office of Science and Office of Basic Energy Sciences of the U.S. Department of Energy under Contract No. DE-AC02-05CH11231.

**Author Contributions**

M.S. and N.A.C. designed the research. M.S., A.K., Y.S., G.P.S., M.R.T., C.Z. and N.A.C. performed the experiments. A.M., D.L. and M.C supplied the materials. M.S., R.G.P., M.A.G., J.E.M. and N.A.C. analyzed and interpreted the data. M.S., J.E.M. and N.A.C. wrote the manuscript. All the authors read and commented on the manuscript.

**Competing financial interests**

The authors declare no competing financial interests.

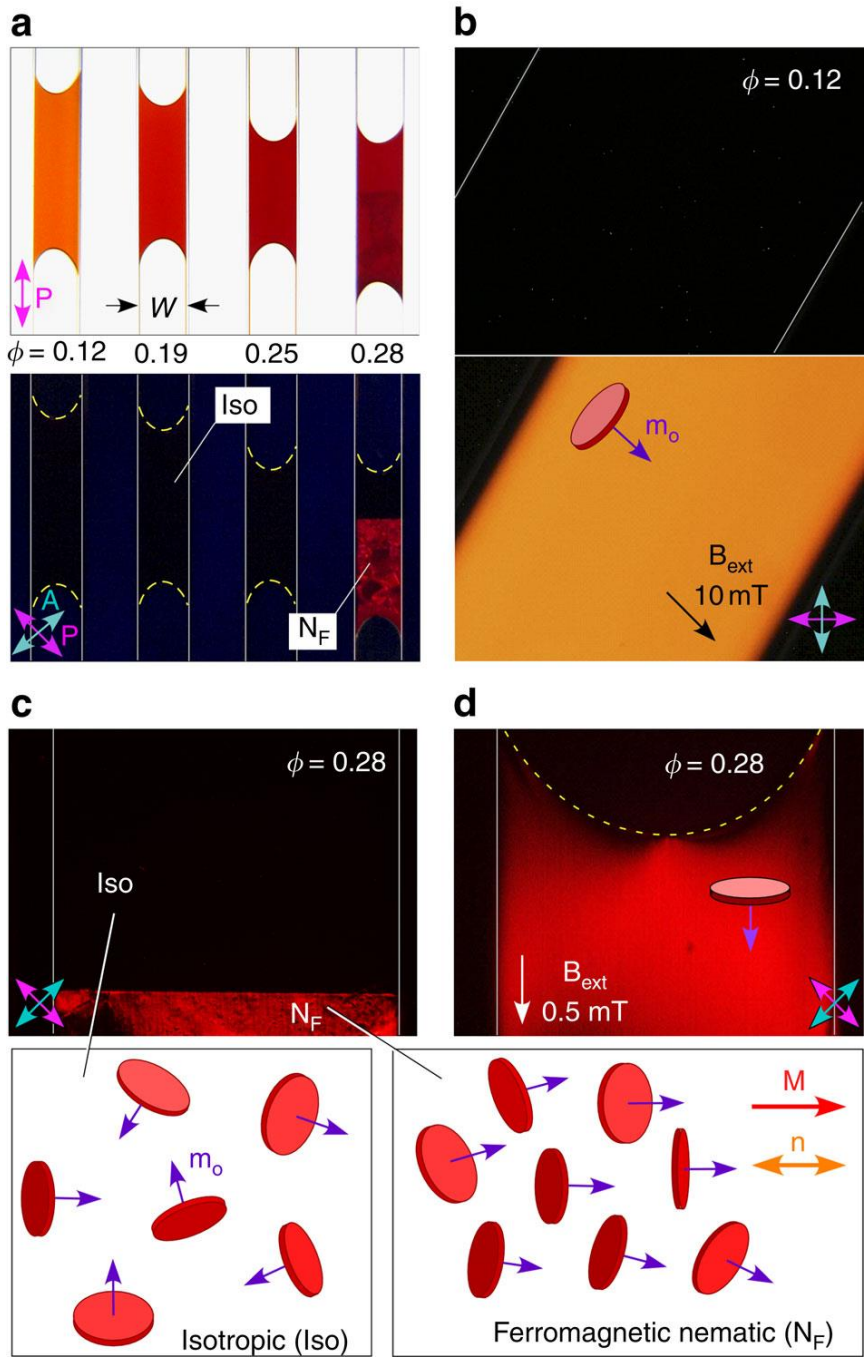

**Figure 1: Liquid crystal ordering of barium ferrite nanoplatelets suspended in n-butanol.**

(**a**) Barium hexaferrite nanoplatelet suspensions viewed in transmitted light with optical polarization conditions indicated (Polarizer: magenta, **P**; and analyzer: cyan, **A**). Low volume fraction suspensions ($\phi \lesssim 0.25$) are isotropic (Iso), appearing dark between crossed polarizers. The orange/red color is due to optical absorption by the nanoplatelets. At higher concentrations ($\phi \gtrsim 0.28$), a birefringent ferromagnetic nematic (N$_F$) phase appears in the lower part of the cell. (**b**) An applied in-plane magnetic field **B**$_{ext}$ induces birefringence in the isotropic phase, with the principal axes of the optical dielectric tensor along and normal to **B**$_{ext}$ and the induced macroscopic magnetization density **M** parallel to **B**$_{ext}$. (**c**) The N$_F$ phase is separated gravitationally from the isotropic region by a sharp, horizontal interface. Equilibrium Iso and N$_F$ structures deduced from birefringence and dichroism measurements are illustrated, where **m**$_o$ is the nanoplate magnetic moment and **n** the director, indicating the local, mean nanoplate normal. (**d**) The Iso phase is magnetized and the Iso/N$_F$ interface becomes continuous under 0.5 mT applied magnetic field. Samples are sealed in rectangular glass capillaries of thickness $L = 50$ μm and width $W = 1$ mm. The boundaries of the cells are indicated by the solid thin white lines. The air-liquid interfaces are indicated by the dashed yellow lines.

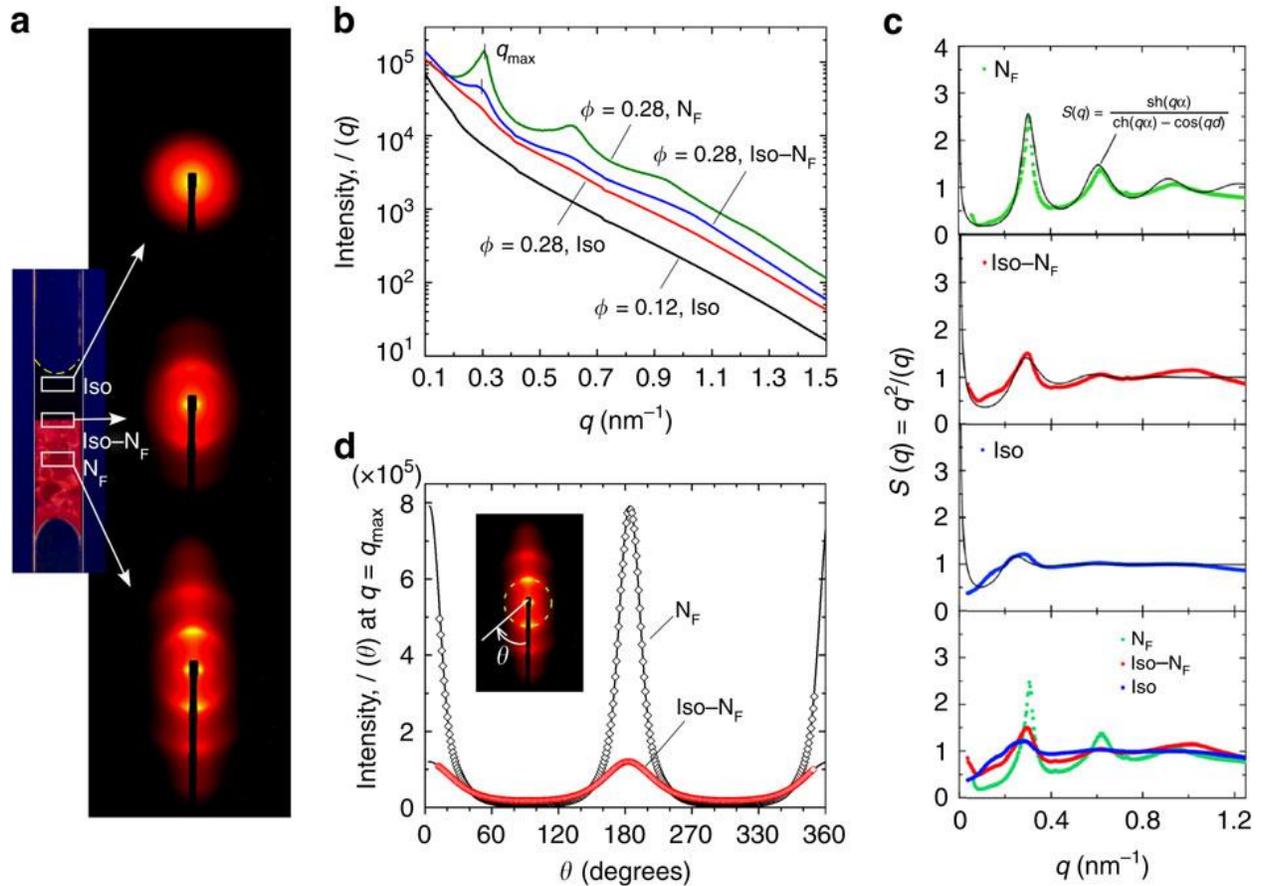

**Figure 2: Synchrotron X-ray diffraction of BF/BuOH magnetic nanoplatelet suspensions.**

A suspension with $\phi = 0.28$ was filled into a flat capillary of thickness $L = 50$ μm and width $W = 1$ mm and sedimented for 1 day, after which it developed an Iso/$N_F$ interface. (**a**) 2D diffraction patterns measured at the three heights in the sample indicated at left by the white boxes representing the beam size and location. *Location Iso*: immediately below the air-liquid interface; *Location Iso/$N_F$:* immediately above the Iso/ $N_F$ interface; *Location $N_F$*: in the $N_F$ phase. (**b**) Diffraction intensity versus $q$ obtained by circular integration of the 2D images in (**a**). The integrated diffraction from a $\phi = 0.12$ sample is shown for comparison. This suspension is uniformly isotropic and the scattering is featureless. (**c**) Structure factor $S(q) = q^2 I(q)$ at the three locations probed in the $\phi = 0.28$ cell. The position of the first diffuse peak moves from $q_{max} = 0.30$ nm$^{-1}$ in the Iso/$N_F$ region to 0.25 nm$^{-1}$ in the Iso region, corresponding to a ~15% density difference between the two locations, consistent with Supplementary Fig. 3. The solid curves are a fit to the equation shown, modeling the layer correlations in a suspension of thin plates, where $\alpha$ is the average of the magnitude of the orientation difference between neighboring nanoplates.

(**d**) Diffraction intensity $I(\theta)$ along the circle passing through the first diffuse peak for the Iso/N$_F$ and N$_F$ locations, respectively. This angular distribution is fitted by the function $I(\theta) = \sum_n p_{2n}\cos^{2n}\theta$ (solid lines), where $\theta$ is the azimuthal angle as illustrated in the inset.

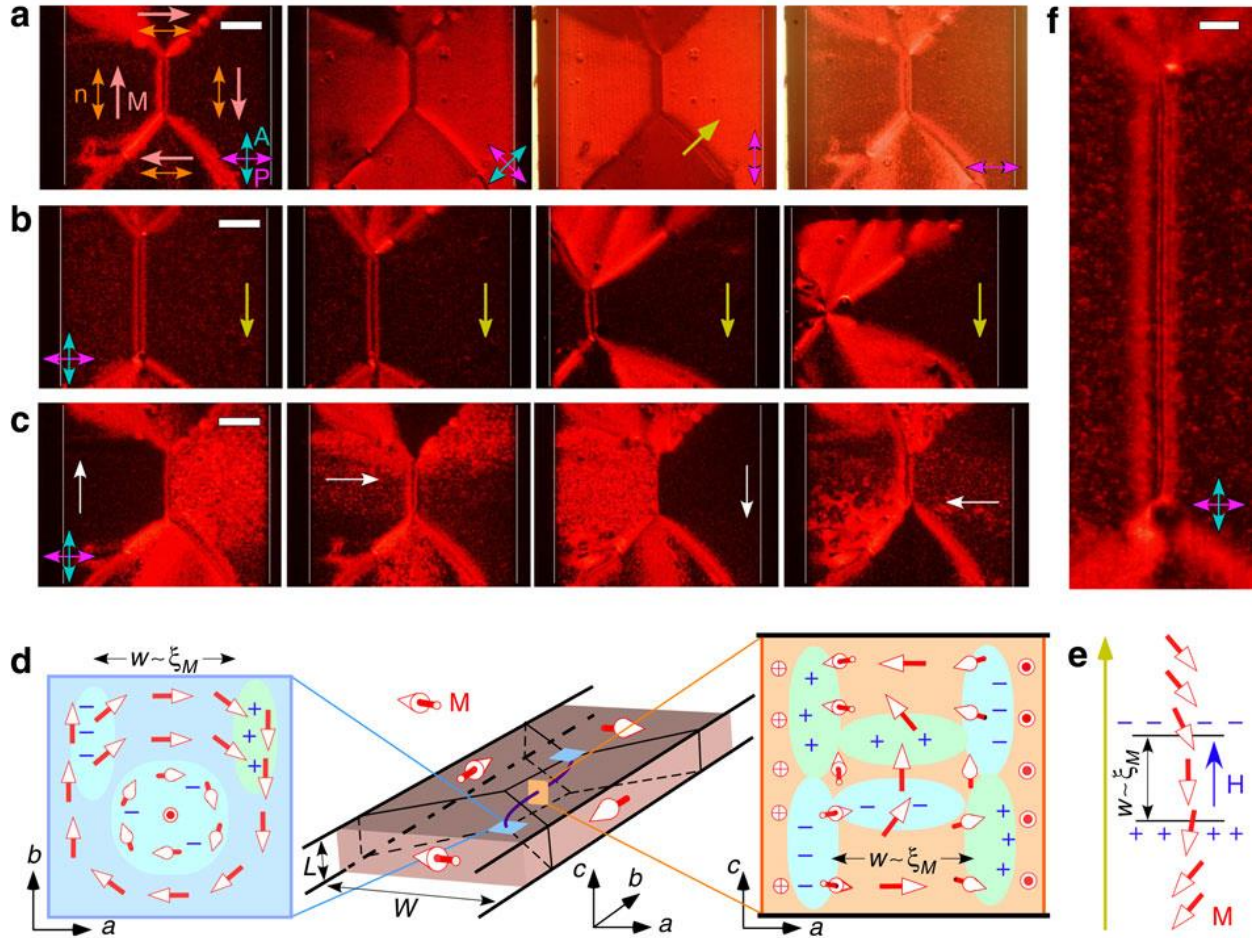

**Figure 3: Equilibrium ferromagnetic nematic flux loop texture in a thin capillary.**

(**a**) Orientational domain textures in the $N_F$ phase in a cell equilibrated in the absence of applied magnetic field viewed in transmission with optical polarization conditions [**P**, **A**] indicated. Uniform domains, with magnetization **M**(**r**) orient in a discontinuous loop to minimize magnetic charge at the domain boundaries. The domains extinguish wherever the director field **n**(**r**) is parallel or normal to crossed polarizers. Rotating the polarizers confirms that the domains are birefringent. Optical dichroism is revealed with **P** only, the nanoplate absorbance being larger for **P** parallel to the nanoplate planes (normal to **n**). (**b**) Applying $B_{equil}$ (yellow arrows; from left: $B_{equil}$=0.03 mT, 0.04 mT, 0.08 mT and 0.10 mT) induces domain wall displacement, allowing the domain with **M** parallel to $B_{equil}$ to grow in area. (**c**) A small transient field (white arrows; $B_{trans}$~0.02 mT) applied parallel to **M** makes the orientation within that domain more uniform (darker). $B_{trans}$ applied antiparallel to **M** produces a random texture of brighter, reorienting subdomains. (**a**–**c**) Scale bar, 200 μm. The cells are 1 mm in width, with the boundaries indicated by thin white lines. (**d**) An illustration of the twist wall passing along the cell centreline from one cell surface to the other. (**e**) Structure of the splay/bend/splay domain boundaries interior to the cell, along the line indicated by the olive arrow in **a**, stabilized by a balance of LC bend elasticity and magnetic charge attraction. (**f**) Magnified image of the +1 escaped twist wall in **n(r)** and **M(r)** imposed by the loop structure of **M(r)**. Scale bar, 50 μm. $L=50$ μm, $W=1$ mm.

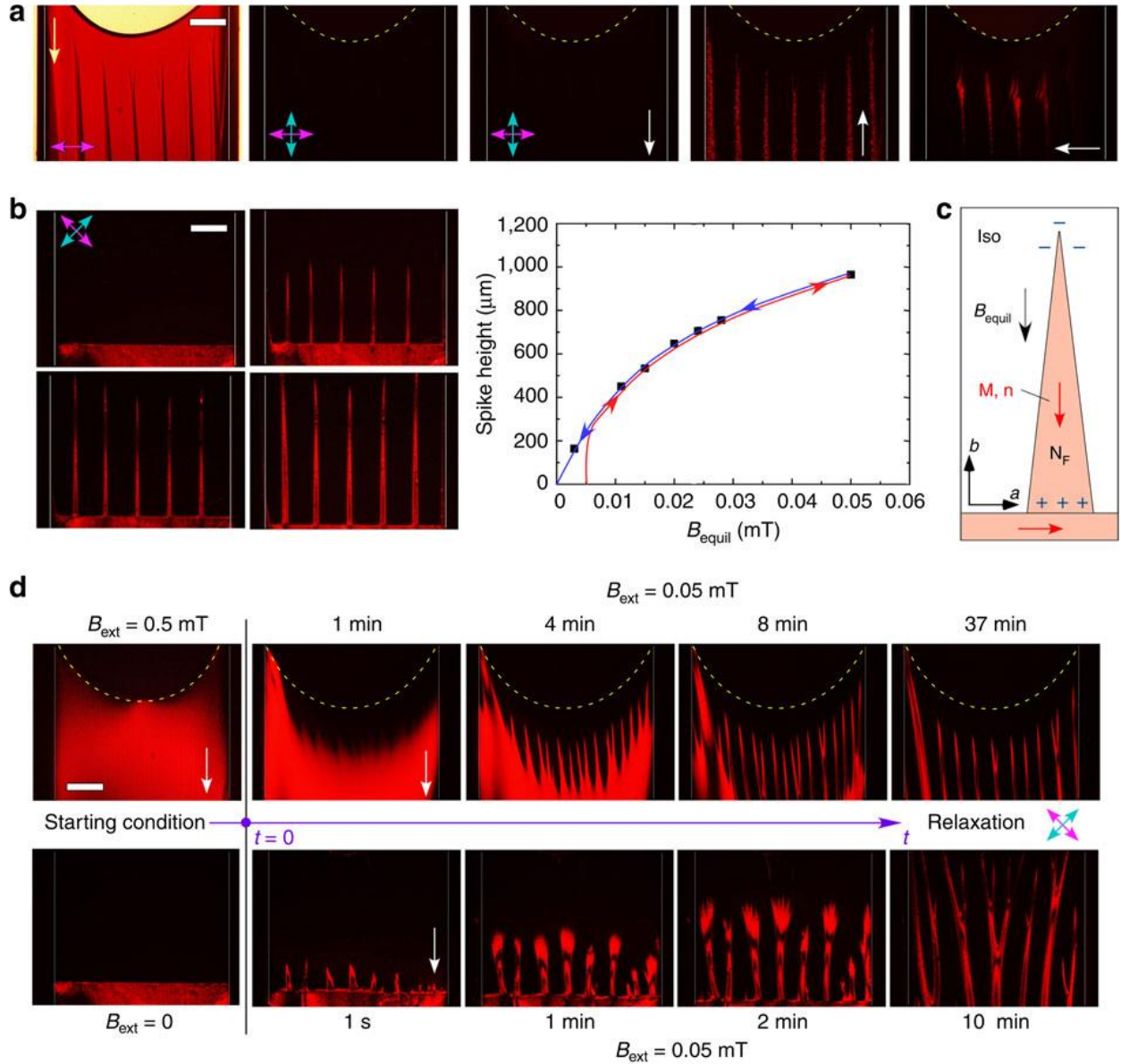

**Figure 4: Ferromagnetic nematic spikes at the BF/BuOH isotropic/nematic interface.**
(**a**) Spikes obtained by equilibrating a $\varphi=0.28$ suspension in a rectangular capillary in an external field (yellow arrow, $B_{equil}=0.05$ mT) for 2 days. The optical polarization conditions [**P**, **A**] are indicated. The reaction of the spikes to transient fields (white arrows, $\mathbf{B}_{trans}\sim 0.02$ mT) indicates that **n** and **M** in the spikes are in the same direction as $\mathbf{B}_{equil}$. (**b**) Spike height versus $B_{equil}$. The spike growth is limited by their penetration into the lower-susceptibility Iso phase in the top of the cell. (**c**) Sketch of ferromagnetic spike at the Iso–$N_F$ phase boundary showing magnetization and magnetic charges. (**d**) Spike dynamics. Application of a large external field ($B_{ext}=0.5$ mT) induces saturated **M**, except near the Iso–air interface. On reducing $B_{ext}$ to 0.05 mT, the $N_F$–Iso interface reforms initially with an undulation instability, which then evolves into spikes. After application of an external field $B_{ext}$=0.05 mT to an initially smooth interface formed at $B_{ext}$=0, on interfaces instability appears and grows into spikes. The cells are 1 mm in width with boundaries indicated by the thin white lines. Scale bar, 200 μm.

SUPPLEMENTARY FIGURES

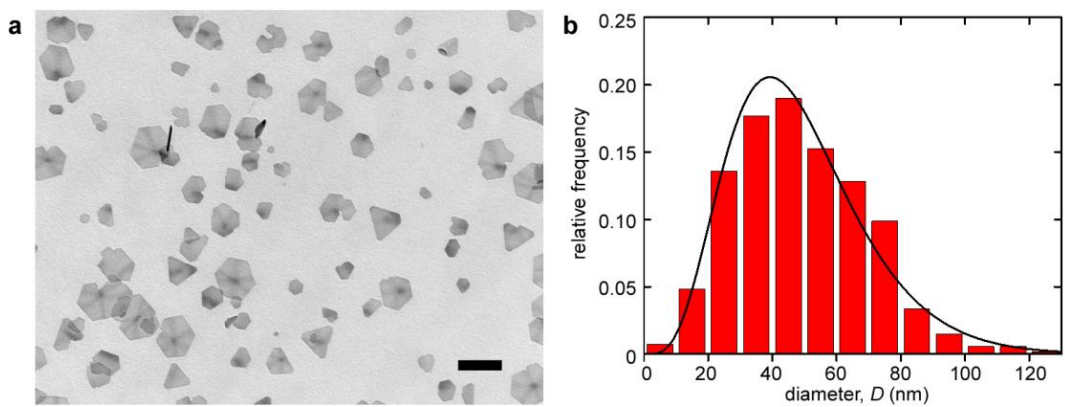

**Supplementary Figure 1: Disk-shaped BF nanoplates and their diameter distribution**.

(**a**) Transmission electron microscope (TEM) image of functionalized BF nanoplates. Scale bar: 100 nm. (**b**) Diameter distribution of the BF nanoploates based on statistical analysis of 500+ nanoplates from TEM images. The measured distribution is fitted by the gamma distribution function, $f(D;\kappa,\tau) = D^{\kappa-1}e^{-D/\tau}/[\tau^{\kappa}\Gamma(\kappa)]$ (solid line), where $\Gamma(\kappa)$ is the gamma function evaluated at $\kappa$, and $\kappa$ and $\tau$ are fitting parameters ($\kappa$ = 5.2 and $\tau$ = 9.2). The mean diameter of the nanoplates is 48 nm with a standard deviation of 21 nm.

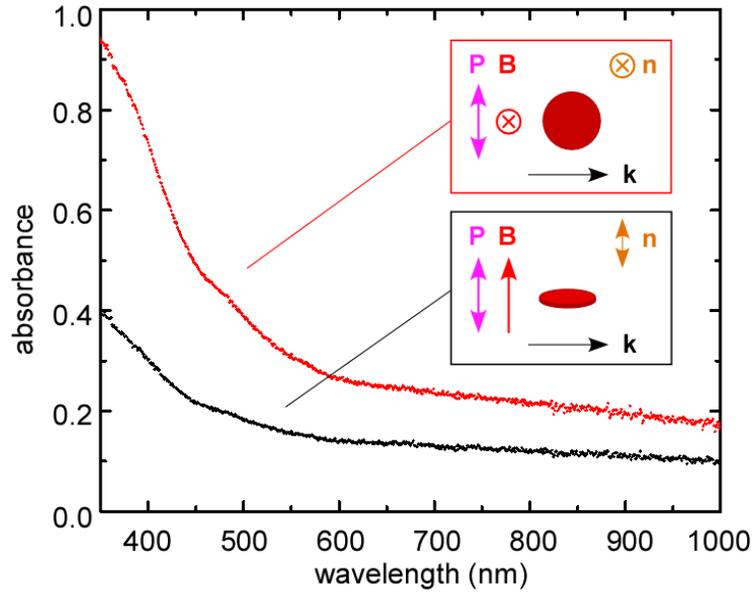

**Supplementary Figure 2: Optical dichroism of disk-shaped BF nanoplates**.

Optical absorbance of barium hexaferrite ($\phi = 0.003$ in 50 µm thick cell) oriented by a 20 mT magnetic field applied either along or perpendicular to the probe beam **k** to induce alignment of the director **n**. The polarization direction **P** of the incident beam is fixed. The BF suspensions absorb strongly at short wavelength, making them appear red. The BF nanoplates are dichroic, absorbing more strongly for polarization in the plane of the nanoplates.

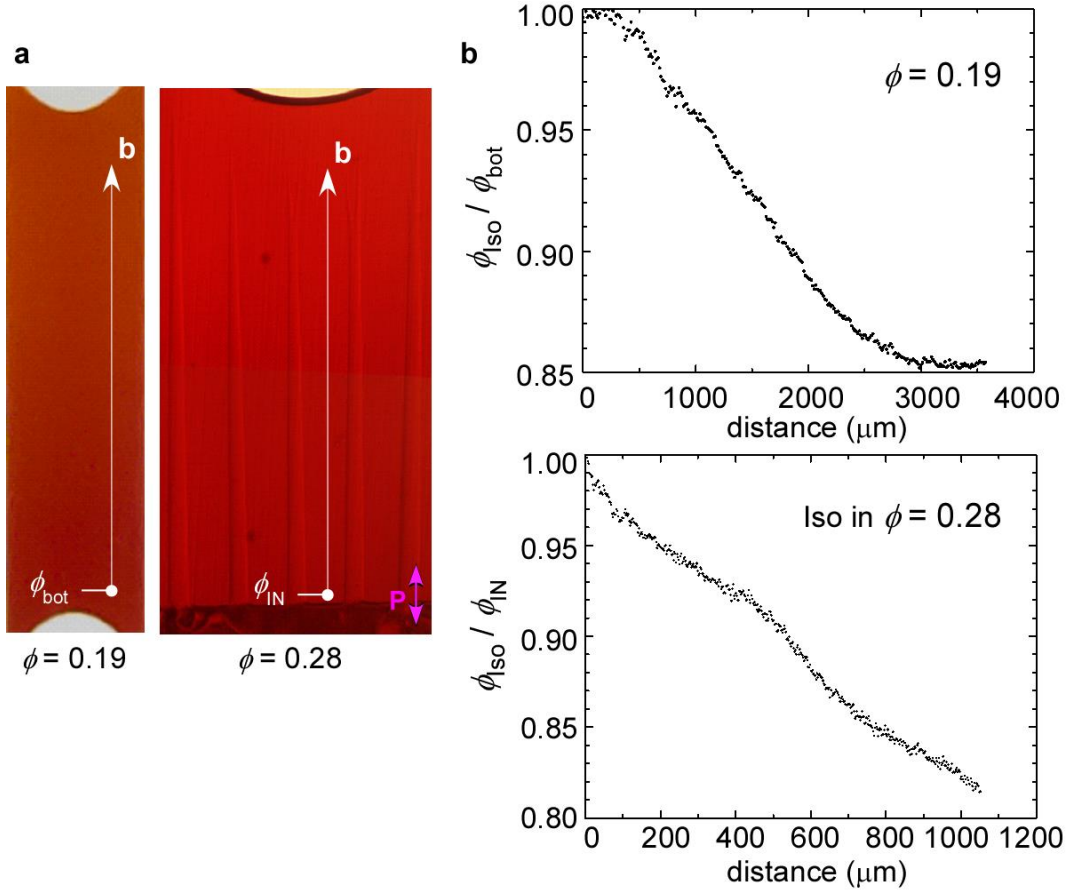

**Supplementary Figure 3: Density gradient of BF nanoplates in the Iso phase**.

(**a**) Typical textures obtained with a single optical polarizer (magenta, **P**) of the $\phi = 0.19$ and $\phi = 0.28$ suspensions equilibrated for two days in a vertical 0.05 mT magnetic field. (**b**) The density – expressed as the relative volume fraction $\phi_{Iso}(b)/\phi_{bot}$ for the $\phi = 0.19$ suspension where $\phi_{bot}$ is the volume fraction at the bottom of the capillary, and $\phi_{Iso}(b)/\phi_{IN}$ for the $\phi = 0.28$ suspension where $\phi_{IN}$ is the volume fraction just above the Iso/$N_F$ interface – is plotted as a function of distance along the **b** axis. In both samples, the gravity-induced concentration gradient in the Iso phase is appreciable but small. Concentration differences of less than 20% were detected over their height of a few millimeters. The relative volume fraction is calculated from the intensity using $\phi_{Iso}(b)/\phi_{IN} = \ln(I_{Iso}(b)/I_o)/\ln(I_{IN}/I_o)$, where $I_o$, $I_{IN}$, and $I_{Iso}(b)$ are respectively the transmitted intensities of the BuOH solvent, of the Iso phase of the BF/BuOH suspension just above the Iso/$N_F$ interface, and of the upper part of the Iso phase.

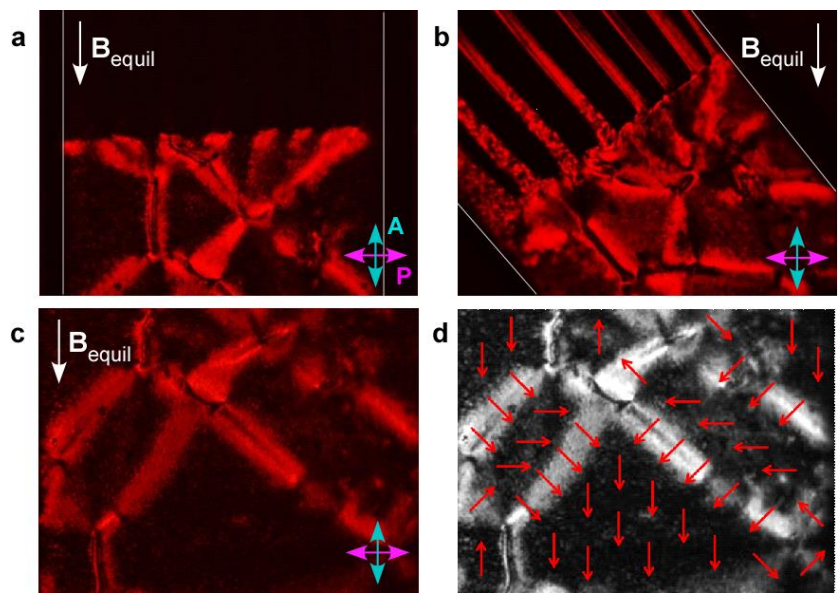

**Supplementary Figure 4: Domain structure and interfacial spikes under 0.05 mT equilibrium magnetic field.**

Capillary BF/BuOH cell at $\phi = 0.28$ with an Iso/$N_F$ interface equilibrated for several days in an external magnetic field $B_{equil} \sim 0.05$ mT, directed downward. (**a**) With the analyzer oriented parallel to $\mathbf{B}_{equil}$, the upper part of the capillary appears dark. (**b**) Rotation of the cell reveals nematic spikes in the Iso phase, which have grown from the interface. The nematic director in the spikes is initially along $\mathbf{B}_{equil}$ but begins to reorient when the cell is rotated, producing the disordered/mottled texture in the leftmost spikes. (**c**) In accordance with Fig. 3b, the block domain with magnetization along the applied downward field has grown to be the largest. (**d**) Magnetization field of the domain structure in (**c**).

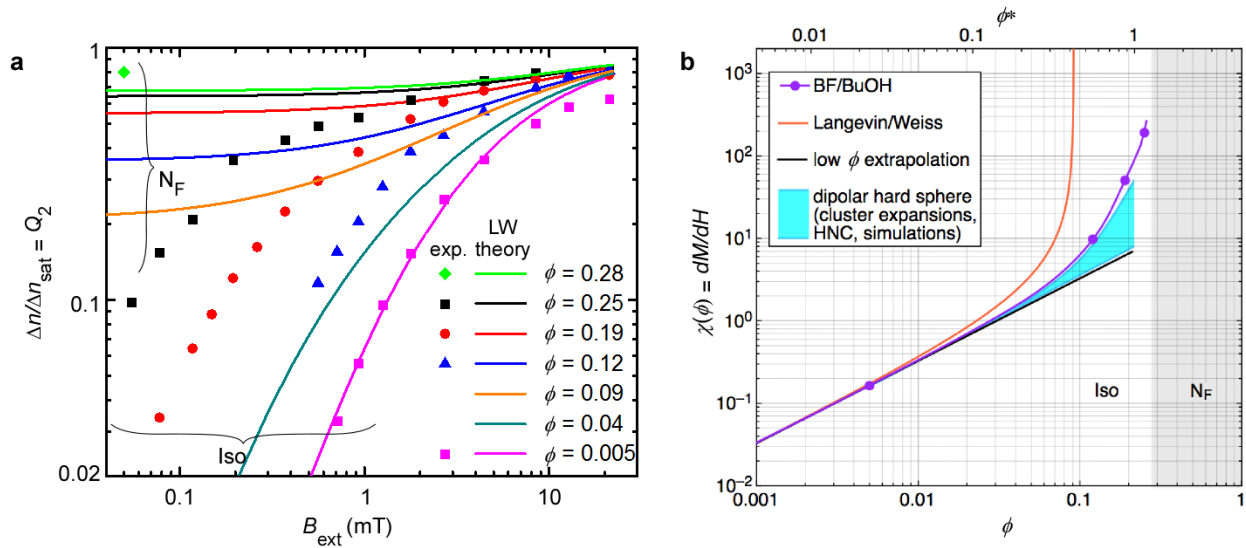

**Supplementary Figure 5: Magnetic field induced birefringence and ordering of the Iso and $N_F$ nanoplate suspensions.**

(**a**) (*symbols*) Experimental birefringence $\Delta n/\Delta n_{sat}$ versus $B_{ext}$ for four BF/BuOH suspensions in the Iso phase at concentrations $\phi$ = 0.25, 0.19, 0.12 and 0.005, and in the $N_F$ phase at $\phi$ = 0.28 and small $B_{ext}$. (*solid lines*) Calculated $\Delta n/\Delta n_{sat}$ versus $B_{ext}$ curves from the LW mean-field theory for polydisperse nanoplate suspensions with volume fractions shown in the legend (Supplementary Eq. 12). The mean magnetic moment $m_o = 2\times10^{-18}$ A m$^2$ is obtained from the best fit to the $\phi$ = 0.005 suspension, giving a dipolar coupling constant $\lambda = \mu_o m_o^2/(4\pi D^3 k_B T) = 0.9$. (**b**) (*symbols*) Experimental values of initial (low field) susceptibility $\chi$ versus $\phi$, derived from data in (**a**) using Supplementary Eqs. 11, 5, and 4 for the four Iso suspensions at concentrations $\phi$ = 0.25, 0.19, 0.12 and 0.005 under low fields. On the top axis indicated is $\phi^* = \nu \pi D^3/6 = 2D\phi/3t$, the effective volume fraction of spheres swept out by rotational diffusion of the discs. The plot shows calculated $\chi$ versus $\phi$ and $\phi^*$ for $\lambda = 0.9$ using several theoretical and simulation methods to account for orientational and positional correlations, as reviewed in Ref. 1. The models all have $\chi = 8\phi^*\lambda$ in the limit of low concentration to match the $\phi$ = 0.005 susceptibility data. The LW mean-field theory then predicts the Iso/$N_F$ transition at a value of $\phi$ that is too small, as is also evident in (**a**). The monodisperse dipolar hard sphere models of Ref. 1 are distributed in the cyan region, with the largest $\chi$ coming from the Reference Limited Hypernetted Chain model of Patey[2], which is also the closest to the experimental data. This supports the notion that at lower concentrations the plates may be approximated as interacting dipolar hard spheres.

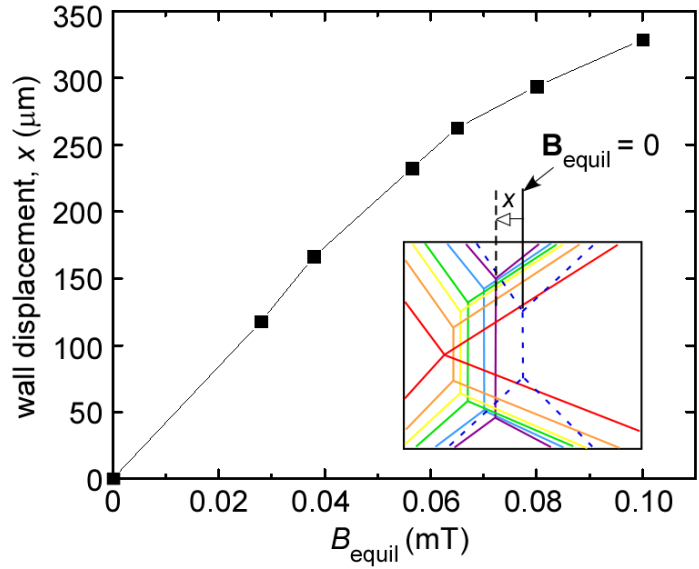

**Supplementary Figure 6: Domain wall displacement under applied magnetic field.**

Neighboring magnetic domains with opposite magnetization directions separated by a domain wall change size as the wall shifts under an applied magnetic field, $B_{equil}$, applied parallel to the magnetization direction of the domain on the right-side of the wall (Fig. 3b). The linear dependence of the position of the wall on field strength indicates an elastic deformation of the wall system in response to field-induced stresses $\sigma_M \propto U_M = -\mathbf{M} \cdot \mathbf{B}_{equil}$. The linear response constitutes compelling evidence for an equilibrium magnetization density in the field-free state. Insert: Domain boundary positions as a function of applied field strength: dashed, $B_{equil} = 0$; purple, $B_{equil} = 0.028$ mT; blue, $B_{equil} = 0.038$ mT; green, $B_{equil} = 0.057$ mT; yellow, $B_{equil} = 0.065$ mT; orange, $B_{equil} = 0.080$ mT; red, $B_{equil} = 0.10$ mT.

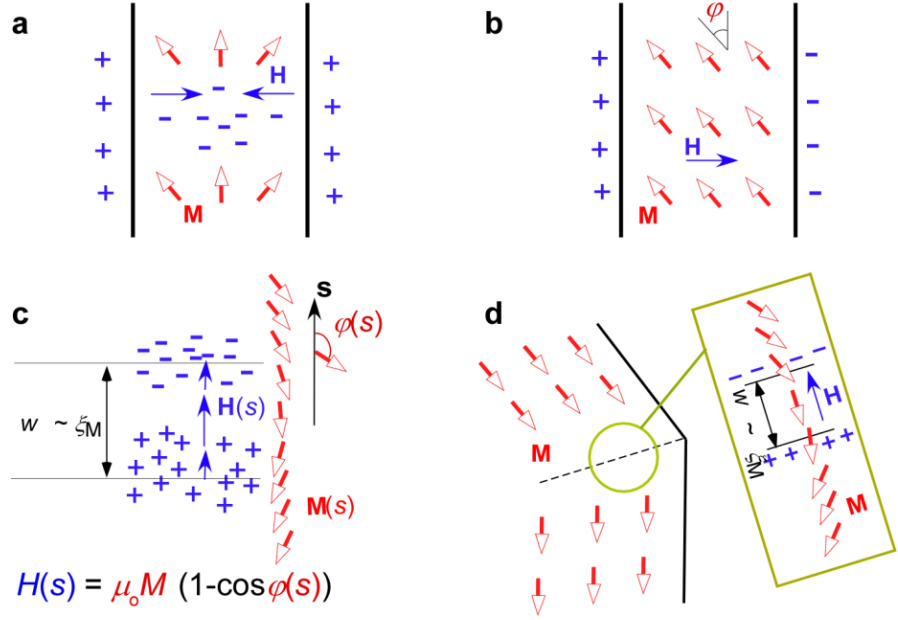

**Supplementary Figure 7: Textures of M(r) at domain boundaries.**

(**a**) Splay distortion of **M**(**r**) between infinite, parallel boundary sheets, showing the associated bulk magnetic charge $\rho_m(\mathbf{r}) = -\nabla \cdot \mathbf{M}(\mathbf{r})$ and surface magnetic charge $\rho_s(\mathbf{r}) = \mathbf{M}(\mathbf{r}) \cdot \mathbf{s}(\mathbf{r})$, which generates an **H** field within the magnetized material. If the splay deformation is reduced to zero, then $H = 0$. (**b**) If **M**(**r**) is uniformly rotated away from being parallel to the infinite, confining boundary sheets by an angle $\varphi$, then the surface magnetic charge generates a uniform field $H = M\sin\varphi$ within the magnetized material. (**c**) Boundaries between two semi-infinite domains of uniform orientation. The boundary conditions cause the planar interface between the two domains to take the form of a two-dimensional splay/bend/splay wall of width $w \sim \xi_M = \sqrt{K/(\mu_o M^2)}$ ($K$ is the LC Frank elastic constant in the one-constant approximation), which is a magnetic counterpart of the case in ferroelectric LCs[3]. $K$ may be taken to be comparable to the splay elastic constant because it is much larger than that for bend. This structure is a soliton-like solution[3] of Supplementary Eq. 20. **s** is the coordinate describing the displacement normal to the wall. (**d**) Structure of a ferromagnetic domain having a planar border with a discontinuous change of direction. The boundary condition prefers **M**(**r**) to be parallel to the border everywhere in order to reduce magnetic charge there. This generates semi-infinite domains of uniform **M**(**r**) that require a combination of splay and bend deformation to connect smoothly. The magnetic energy associated with splay-generated magnetic charge is minimized here by making the region of deformation as thin as possible. The splay/bend/splay structure of (**c**) is the local solution to this condition, creating a domain boundary that has the same local structure everywhere along a line that bisects the angle made by the kink in the border.

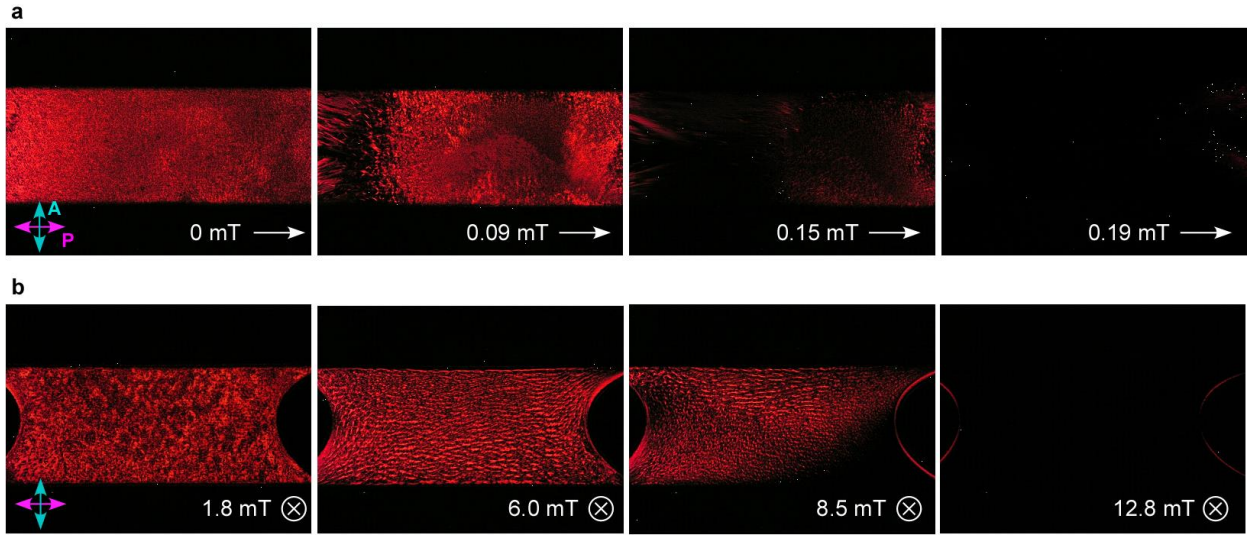

**Supplementary Figure 8: Anisotropic alignment of BF/BuOH in external magnetic fields.**

(**a**) $\mathbf{B}_{ext}$ parallel to the cell plane. An in-plane field $\mathbf{B}_{ext} \sim 0.2$ mT aligns a homogenized $\phi = 0.28$ sample to a bar-magnet-like monodomain. If this field is then reduced, the resulting magnetic charge on the poles of the bar magnet generate a demagnetizing field that reorients the LC into an array of small, randomly oriented domains, just as shown in the first image on the left. (**b**) $\mathbf{B}_{ext}$ normal to the cell plane. Substantial reorientation of **M** and **n** out of the cell plane requires a large field, $\mathbf{B}_{ext} \gtrsim 13$ mT.

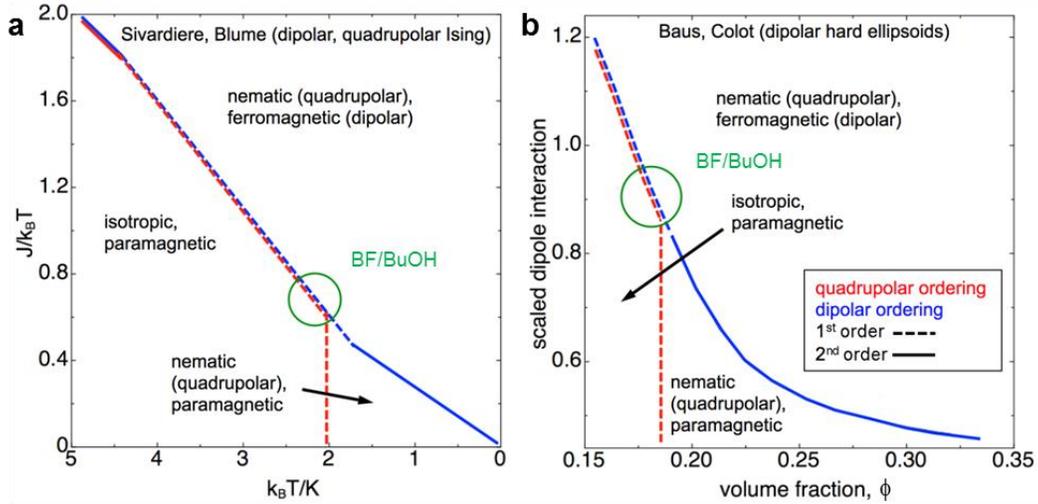

**Supplementary Figure 9: Mean-field phase diagrams of model systems with dipolar and quadrupolar interactions.**

(**a**) Sivardiere/Blume Ising model on a lattice with a Hamiltonian of nearest-neighbor dipolar and quadrupolar interactions[4]. $J$ is the pair interaction energy of the dipolar term and $K$ the interaction energy of the quadrupolar term. The phase diagram is calculated in the molecular field approximation. (**b**) Baus/Colot off-lattice model of hard ellipsoids of revolution with embedded dipoles[5]. $\phi$ is the volume fraction of the ellipsoids.

**Supplementary Note 1: Field-induced magnetization and birefringence in the Iso phase**

In order to assess the nature of the magnetic ordering in the Iso phase, $\Delta n(B_{\text{ext}})/\Delta n_{\text{sat}}$, where $\Delta n_{\text{sat}}$ is the limiting $\Delta n$ at high fields, is measured for several BF/BuOH concentrations, using 630 nm monochromatic light. The results are shown in Supplementary Fig. 5a. To analyze the data, we first tried to apply the standard Langevin-Weiss (LW) mean-field model[6], accounting for the polydispersity of the particle sizes, which enables calculation of $Q_1$ and $Q_2$ for a system of magnetic dipoles in thermal equilibrium. For a given diameter distribution function, the free parameters in the model are the temperature $T$ (here always 298 K), the total particle number density $\nu$, and the mean nanoplate magnetic moment $m_o$. For the purpose of this calculation, the nanoplates are divided into $i = 23$ species according to their diameters with a 5 nm bin size and low and high end cut-offs at 10 nm and 120 nm, respectively, following the fitted distribution function $f(D; \kappa, \tau) = D^{\kappa-1} e^{-D/\tau} / [\tau^{\kappa} \Gamma(\kappa)]$, where $\kappa = 5.2$ and $\tau = 9.2$. In the absence of an applied magnetic field, the nanoplate moments in the Iso suspensions of orient randomly. When the Iso suspensions are subjected to an applied field, the magnetic torques interaction tend to align the magnetic moments of individual nanoplates with the field. If the field is strong enough, the magnetic moments of all the nanoplates will be completely aligned and the magnetization of the suspension will reach the saturation value $M_{\text{sat}}$

$$M_{\text{sat}} = \langle \nu m_o \rangle = \sum_i \nu_i m_i , \qquad (1)$$

where the subscribed $i$ represents the $i$th species.

When the magnetic field is not strong enough to completely align the magnetic moments of the nanoplates, thermal fluctuations tend to randomize the orientation of the nanoplates. The effective magnetic moment of a nanoplate is its component along the field direction, $m_i \cos \vartheta_i$, where $\vartheta_i$ is the angle between $\mathbf{m}_i$ and the applied magnetic field $\mathbf{B}_{\text{ext}}$. The field-induced magnetization of the suspension is the total effective magnetic moment per unit volume. It can be described by the first-rank order parameter $Q_{1i}(B_{\text{ext}})$

$$Q_{1i}(B_{\text{ext}}) = \langle \cos \vartheta_i \rangle = \frac{\int_0^{\pi/2} g(\vartheta_i) \cos \vartheta_i \sin \vartheta_i d\vartheta_i}{\int_0^{\pi/2} g(\vartheta_i) \sin \vartheta_i d\vartheta_i}, \tag{2}$$

with $g(\vartheta_i)$ being the orientational distribution function for the $i$th species. $g(\vartheta_i)$ can be expressed as

$$g(\vartheta_i) = \exp(m_i B \cos \vartheta_i / k_B T), \tag{3}$$

where $k_B$ is the Boltzmann constant and $B$ is the total magnetic field. For convenience, we define

$$\gamma_i = m_i B / k_B T. \tag{4}$$

After integration, we obtain the so-called Langevin function

$$Q_{1i}(B) = \langle \cos \vartheta_i \rangle = \coth \gamma_i - \frac{1}{\gamma_i}. \tag{5}$$

The total magnetic field is a sum of the applied magnetic field $B_{\text{ext}}$ and the internal magnetic field from the magnetized suspension

$$B = B_{\text{ext}} + \frac{1}{3} \mu_o \sum_i M_i, \tag{6}$$

where the second term describes the local magnetic field on a nanoplate in a spherical cavity of surrounding nanoplates. Taking all the species in a suspension into consideration, the overall first-rank order parameter is then

$$Q_1 = \frac{\langle M \rangle}{\langle v m_o \rangle} = \frac{\sum_i M_i}{\sum_i v_i m_i}. \tag{7}$$

The field-induced birefringence, $\Delta n(B_{\text{ext}})$, can be expressed in terms of the second-rank order parameter $Q_2$ as

$$\Delta n_i = \alpha \pi v_i D_i^2 Q_{2i} / (8 n_m), \tag{8}$$

where $\alpha$ is the optical polarizability anisotropy per unit area of the nanoplates, $n_m$ is the mean refractive index of the solution, and

$$Q_{2i} = \left\langle \frac{3 \cos^2 \vartheta_i - 1}{2} \right\rangle. \tag{9}$$

The mean value of $\cos^2 \vartheta_i$ is obtained by integrating over the entire distribution

$$\left\langle \cos^2 \vartheta_i \right\rangle = \frac{\int_0^\pi g_i(\vartheta_i) \cos^2 \vartheta_i \sin \vartheta_i d\vartheta_i}{\int_0^\pi g_i(\vartheta_i) \sin \vartheta_i d\vartheta_i}, \quad (10)$$

which yields

$$Q_{2i} = 1 - \frac{3}{\gamma_i} \coth \gamma_i + \frac{3}{\gamma_i^2}. \quad (11)$$

When the field is strong enough that all the nanoplates are perfectly aligned, $Q_{2i} = 1$ and $\Delta n = \Delta n_{\text{sat}}$. Taking all the species in the suspension into consideration, the overall second-rank order parameter is

$$Q_2 = \frac{\Delta n}{\Delta n_{\text{sat}}} = \frac{\sum_i v_i D_i^2 Q_{2i}}{\sum_i v_i D_i^2}. \quad (12)$$

For low concentration suspensions, the magnetic coupling between nanoplates should be weak and the LW theory a reasonable approximation. For a $\phi = 0.005$ suspension, we find that $m_o = 2 \times 10^{-18}$ A m$^2$. This is very close to the value of $3 \times 10^{-18}$ A m$^2$ measured[7] for aligned and dried colloidal nanoplates with a mean diameter of 70 nm.

**Supplementary Note 2: Low-field susceptibility of the Iso phase**

To further understand the magnetization process in the nanoplate suspension system, we measured the field-induced birefringence $\Delta n(B_{\text{ext}})/\Delta n_{\text{sat}}$ versus $B_{\text{ext}}$ in Iso suspensions of different concentration $\phi$, shown in Supplementary Fig. 5a. From these data we extracted the initial (low-field) magnetic susceptibility, $\chi = dM/dH$, using the response at low fields along with Supplementary Eqs. 11, 5 and 4, under the assumption of monodisperse particles. Supplementary Fig. 5b shows the resulting $\chi$ (symbols) and the prediction of the LW mean field model. Both Supplementary Figs. 5a and b indicate that if the LW model is adjusted to fit $\chi$ in the limit of low $\phi$, then the predicted Iso/N$_F$ transition occurs at a much lower concentration than observed experimentally.

Since here the nanoplates are in the Iso phase, we may consider them to be discs with orientational diffusion, sweeping out volumes obtained by spinning the average size disc about its diameter and in doing so behaving like hard spheres, as in an Onsager isotropic phase. Pursuing this notion, we indicate in Supplementary Fig. 5b the range of susceptibilities obtained from several theoretical models for the susceptibility of monodisperse dipolar hard spheres[1], which account for the dipolar coupling and correlation in different ways. The equivalent spherical volume fraction $\phi^*$ is related to the nanoplate volume fraction $\phi$ by $\phi^* = \nu\pi D^3/6 = 2D\phi/3t$. These models all behave at low $\phi$ as the same distribution of independent plates, giving the same $\chi(\phi) \propto \phi$, and are therefore all scaled in the same way to match the experiment at low $\phi$ by setting $m_o$ to $2\times10^{-18}$ A m$^2$. The dipolar coupling constant[1] used for the dipolar hard spheres curves in Supplementary Fig. 5b is $\lambda = \mu_o m_o^2/(4\pi D^3 k_B T) = 0.9$.

The range of susceptibilities obtained from the dipolar hard sphere models in Fig. 4 of Ref. 1 are indicated by the cyan-shaded region in Supplementary Fig. 5b. At the low limit of the cyan susceptibility range is the Onsager approximation for dipolar spheres[8], and at the high limit the Reference Limited Hypernetted Chain model by Patey[2], the latter providing a reasonable qualitative description of $\chi(\phi)$. The concentration variable $\phi^*$ is that typically used in models of liquid crystal ordering by the steric interaction of discs, being defined such that $\phi^* \sim 1$ at the concentration where the effective spherical volumes occupied by nanoplates start to overlap each other, which is where the Onsager LC ordering of discs occurs. Thus, as $\phi^*$ approaches 1, $\chi(\phi)$ increases faster than the dipolar hard sphere models, which we interpret to be due to enhanced correlations between the discs coming from their flat shape, beyond that arising from the magnetic interaction. Note that for monodisperse hard spheres, the value of $\phi^*$ can never exceed ~0.74, which is the upper limit of the packing fraction of spheres. However, for the case of our nanoplates, $\phi^* = 1$ corresponds to a nanoplate volume fraction of only $\phi \sim 0.22$, and above this concentration is where the Onsager excluded volume theory predicts that an Iso to nematic phase transition occurs[9]. Numerical simulations of polydisperse and charged disc suspensions[10-12] show that an Iso/Nematic transition starts to take place at $\nu\langle D^3\rangle = 3.2 \sim 4.0$. Considering that the volume of a nanoplate is $\pi\langle D^2 t\rangle/4$, and assuming the size distribution shown in Supplementary Fig. 1b, we can calculate the equivalent Iso/N$_F$ phase transition concentration $\phi = 0.27 \sim 0.34$, which corresponds very well with our experimental observations. Thus, at $\phi \sim 0.28$ the BF/BuOH system is near an Onsager-type nematic ordering transition that enhances translational entropy by orienting hard discs to reduce their mutual excluded volume. This nematic ordering transition leads to the dramatic increase of $\chi$ and to the formation of the N$_F$ phase.

**Supplementary Note 3: Orientational order parameters of $N_F$ BF/BuOH suspensions from X-ray diffraction**

The birefringence and dichroism of the domains in the $N_F$ phase indicate orientational ordering of the nanoplate planes. The order parameter of this uniaxial $N_F$ phase can be calculated from these intensity distribution functions as follows. The orientational order of plates in the $N_F$ phase is described by an orientational distribution function $f(\beta)$, in which $\beta$ is the angle between the axis of the plates and the macroscopic symmetry axis of the $N_F$ phase, similar to studies on thermotropic liquid crystals with molecules that possess the cylindrical symmetry. Following the method first established by Leadbetter and Norris[13], the scattered intensities from a uniaxially aligned liquid crystal depends on the orientational distribution as follows:

$$I(\theta) = \int_{\beta=\theta}^{\pi/2} f(\beta) \frac{\sin\beta \sec^2\theta}{\sqrt{\tan^2\beta - \tan^2\theta}} d\beta, \quad (13)$$

where $I(\theta)$ is the azimuthal profile, as illustrated in Fig. 2d (insert). To find $f(\beta)$ from the above function, we follow the method developed by Davidson *et al* [14]. First, $f(\beta)$ is expanded as a Fourier series as follows

$$f(\beta) = \sum_{n=0}^{\infty} f_{2n} \cos^{2n}\beta. \quad (14)$$

By inserting this in Supplementary Eq. 13, we obtain, after integration,

$$I(\theta) = \sum_{n=0}^{\infty} f_{2n} \frac{2^n n!}{(2n+1)!!} \cos^{2n}\theta. \quad (15)$$

The nematic order can be described by the second-rank order parameter

$$Q_2 = \left\langle \frac{3\cos^2\beta - 1}{2} \right\rangle, \quad (16)$$

where

$$\langle \cos^2\beta \rangle = \frac{\int_0^{\pi/2} f(\beta)\cos^2\beta \sin\beta d\beta}{\int_0^{\pi/2} f(\beta)\sin\beta d\beta}. \quad (17)$$

Applying Supplementary Eq. 14, we obtain

$$\left\langle \cos^2 \beta \right\rangle = \frac{\sum_{n=0}^{\infty} \frac{f_{2n}}{2n+3}}{\sum_{n=0}^{\infty} \frac{f_{2n}}{2n+1}}, \tag{18}$$

from which we may obtain the order parameter. $Q_2 = 1$ represents a perfectly oriented state, while $Q_2 = 0$ describes an isotropic state. It is found that $Q_2 = 0.8$ at *Location $N_F$* and $Q_2 = 0.4$ at *Location Iso/$N_F$*. The form factor of the nanoplates, which broadens the distribution by the ratio $(2/D)/(2\pi/d) \sim 7°$, was ignored in this estimate.

**Supplementary Note 4: Ferromagnetic nematic magneto-elastic deformation energy**

We analyze the textural features of the ferromagnetic nematic phase as phenomena resulting from the combined effects of magnetostatic and Frank nematic elastic energies, described by

$$U = \frac{\mu_o M^2}{2} \int d\mathbf{r} d\mathbf{r}' \, [\nabla \cdot \mathbf{n}(\mathbf{r}) \nabla \cdot \mathbf{n}(\mathbf{r}') + (\mathbf{n}(\mathbf{r}) \cdot \mathbf{s}(\mathbf{r_s}) \delta(\mathbf{r} - \mathbf{r_s}))(\mathbf{n}(\mathbf{r}') \cdot \mathbf{s}(\mathbf{r_s}') \delta(\mathbf{r}' - \mathbf{r_s}'))] |\mathbf{r} - \mathbf{r}'|^{-1}$$
$$+ \int d\mathbf{r} \tfrac{1}{2} [K_S (\nabla \cdot \mathbf{n}(\mathbf{r}))^2 + K_T (\mathbf{n}(\mathbf{r}) \cdot \nabla \times \mathbf{n}(\mathbf{r}))^2 + K_B (\mathbf{n}(\mathbf{r}) \times \nabla \times \mathbf{n}(\mathbf{r}))^2] + \int d\mathbf{r} \mathbf{M}(\mathbf{r}) \cdot \mathbf{B}(\mathbf{r})_{\text{ext}}, \tag{19}$$

We assume here that the magnetic moments $\mathbf{m}_o$ are fixed to be normal to the colloidal plates, making $\mathbf{M}(\mathbf{r})$ and $\mathbf{n}(\mathbf{r})$ locally parallel everywhere, *i.e.*, $\mathbf{M}(\mathbf{r}) = M\mathbf{n}(\mathbf{r})$. The nematic Frank elastic energy includes the usually splay, bend, and twist deformation terms. The self-interaction of the magnetic dipole field has been written in terms of the bulk and surface magnetization charge densities $\rho_m(\mathbf{r}) = -\nabla \cdot \mathbf{M}(\mathbf{r}) = -M \nabla \cdot \mathbf{n}(\mathbf{r})$ and $\rho_s(\mathbf{r_s}) = M\mathbf{n}(\mathbf{r}) \cdot \mathbf{s}(\mathbf{r_s}) \delta(\mathbf{r}-\mathbf{r_s})$, respectively.

As discussed in the main text, strong magnetic charge effects confine $\mathbf{M}(\mathbf{r})$ in the sample of Fig. 3 to be parallel to the **a,b** plane, with the texture corresponding to a rotation of $\mathbf{M}(\mathbf{r})$ about **c**. In the case that the characteristic length $\xi_M = \sqrt{K/(\mu_o M^2)}$ is small compared to the sample thickness $L$, we can take the sample to be infinite in thickness and uniform along **c**, giving an orientation field $\varphi$ that is 2D. For an in-plane sinusoidal reorientation of $\mathbf{M}(\mathbf{r})$ of wavevector **q**, Supplementary Eq. 19 then yields a magnetic space charge modulation giving an energy/volume

$$U_\mathbf{q} = \tfrac{1}{2} [\mu_o M^2 (\cos\psi)^2 + K_B (q\sin\psi)^2 + K_S (q\cos\psi)^2] \, |\delta\varphi_\mathbf{q}|^2, \tag{20}$$

where $\psi$ is the angle between **q** and $\delta\mathbf{n} = \mathbf{c} \times \mathbf{n}$, showing explicitly the dominance of magnetostatic splay interactions at long length scale (small $q$). This energy exhibits a wavevector dependence such that magnetostatic energy dominates Frank elasticity and suppresses splay deformation ($\mathbf{q} \perp \mathbf{M},\mathbf{n}$) on length scales longer than $\xi_M$. At shorter length scales, the LC can resist the magnetic torques. This behavior can be seen explicitly for the 3D sample with $\mathbf{M}(\mathbf{r})$ in the **a,b** plane and exhibiting a 2D rotation field $\varphi$ about **c**, by using the applicable torque balance equation coming from Supplementary Eq. 19. For the splay/bend/splay wall shown in Fig. 3e and Supplementary Fig. 7c,d, spatial dependence of $\varphi(s)$ along a single coordinate **s** describing the displacement normal to the wall, we have

$$K \frac{\partial^2 \varphi}{\partial s^2} = \mu_o \mathbf{M} \times \mathbf{H} = \mu_o M^2 \sin\varphi(s)[\cos\varphi_o - \cos\varphi(s)], \tag{21}$$

where we have taken $K_S = K_B = K$, **H** is generated by the magnetic charge, and $\varphi$ is in the range $\varphi_o < \varphi < \varphi_o + \pi$. The soliton-like analytic solution for the wall structure is sketched in Supplementary Fig. 7c, a result initially obtained for the analogous electric case[3]. Assuming $K$ the value of the splay constant $K_S = 6k_BT/D = 5 \times 10^{-13}$ N, obtained from Monte Carlo simulation of cut-spheres[15] with thickness-over-diameter ratio of 1/10 and $Q_2 \sim 0.8$, we find $\xi_M \sim 0.1$ µm. Nematic regions substantially larger than this, such as those shown in Fig. 3, are thus expected to have uniform orientation of $\mathbf{M}(\mathbf{r})$ and $\mathbf{n}(\mathbf{r})$ [16].

In the other limiting case, of a thin sample that is uniform along **c**, with spatial variation $\varphi$ in the **a,b** plane, but where $\xi_M$ is larger than with the sample thickness $L$, the normal mode energy/area becomes:

$$U_\mathbf{q} = \frac{1}{2}[\mu_o(ML)^2|q\cos\psi| + K_BL(q\sin\psi)^2 + K_SL(q\cos\psi)^2] \, |\delta\varphi_\mathbf{q}|^2. \tag{22}$$

The crossover length in this case is $\xi_{ML} = K/(\mu_o M^2 L)$, analogous to the result obtained for freely suspended ferroelectric LC films[16-18].

Analysis based on Supplementary Eq. 19 shows that for a slab of infinite area, the magnetic energy of a uniform **M** field free to orient in any direction is lowest when **M** is parallel to the plane of the slab, in which case: (*i*) there is a uniform magnetic field $\mathbf{B}_M = \mu_o\mathbf{M}$ everywhere in the slob, and the macroscopic mean field in the magnetized material, oriented parallel to and stabilizing **M**, yields the lowest achievable magnetic energy ($U_M = -\mu_o M^2$). An estimate of $\mathbf{B}_M$, assuming $m_o = 2\times10^{-18}$ A m$^2$, is $B_M = \mu_o M \sim \mu_o Q_1 \Sigma v_i m_i \sim 3\times10^4 \mu_o$ A m$^{-1} \sim 40$ mT; (*ii*) $B = 0$ outside of the slab; and (*iii*) the magnetic energy increases harmonically as $\delta U_M(\varphi) = \mu_o M^2 \varphi^2/2$ for rotation $\varphi$ of $\mathbf{M}$ out of the cell plane. An external field $\mathbf{B}_{\text{ext}\perp}$ applied normal to the cell plane couples to the magneticzation with an energy density $U_M(\varphi) = -MB_{\text{ext}\perp}\varphi$ and leads to a field-induced rotation of $M$, $\varphi(B_{\text{ext}\perp}) = B_{\text{ext}\perp}/\mu_o M$, minimizing the free energy density with regard to $\varphi$.

**Supplementary Note 5: Comparison of textural behavior in the BF/BuOH and BF/5CB systems**

Comparison of the relative magnitudes of the energy terms in Supplementary Eq. 20 for the BF/BuOH and BF/5CB nematics[7] shows that the magnetic dipole interaction ($M^2$) term is larger by a factor of ~$10^5$ in BF/BuOH because of the larger density of magnetic nanoplates. At the same time, the Frank elastic constants in BF/BuOH are smaller ($K_B \sim 1\times10^{-13}$ N, $K_S \sim 5\times10^{-13}$ N), relative to those of BF/5CB ($K_B$, $K_S \sim 5\times10^{-12}$ N). These ratios combine to make the characteristic length $\xi_M = \sqrt{K/(\mu_o M^2)}$, expressing the balance between Frank elastic and magnetostatic torques, much smaller than the cell thickness. The result is an array of large local domains, separated by walls of dimension $\xi_M$ that can respond independently to applied field. If the field is applied transverse to $\mathbf{M}$, a uniform reorientation results. However, if, as is the case of Fig. 3c, the field is applied in the –$\mathbf{M}$ direction, depending on their initial local tilt, the domains reorient in random directions, which is a distinctive feature of the regime where $\xi_M$ is much smaller than the sample thickness.

In the BF/5CB case, the length $\xi_M$ is much larger than the cell thickness, meaning that the $M^2$ self-interaction is irrelevant and the field response is determined by the balance of Frank elasticity and the interaction with $B_{\text{ext}}$, yielding, for example, a Fréedericksz transition if $B_{\text{ext}}$ is applied antiparallel to $\mathbf{M}$, with a threshold where $B_{\text{ext}}$ is large enough to reduce the magnetic coherence length $\xi_B = \sqrt{K/(MB_{\text{ext}})}$ to the cell thickness.

**Supplementary Note 6: Analogy to high-polarization ferroelectric liquid crystals**

The scenario of uniform blocks of magnetization separated by sharp domain walls is an example of "orientational fracture", similar to that found[17] in -1 topological defects in the director field structure and textures of thermotropic ferroelectric smectic C liquid crystals with large permanent polarization $P$. In this case, the polarization charge $\rho_p(\mathbf{r}) = -\nabla \cdot \mathbf{P}(\mathbf{r})$ produced by splay deformation of the polarization density $\mathbf{P}(\mathbf{r})$ field is sufficiently costly in energy that splay of $\mathbf{P}(\mathbf{r})$ is expelled from the bulk of a texture, rather being confined to narrow, $\mathbf{P}$-stabilized 1/2 rotation walls of width $w \sim \xi_E = \sqrt{K_S/(\varepsilon_o P^2)}$ separating "block" domains of uniform $\mathbf{P}(\mathbf{r})$. Elastic energy functions of the form of Supplementary Eq. 19 have been employed extensively to describe such behavior in the absence of bulk free-charge screening of the polarization charge[19]. In the electric case, observation of the self-interaction effects of $\mathbf{P}$ require the polarization to be large enough to overcome the screening effects of free charge, from ions in solution, for example. Free charge on surfaces can also control the orientation of $\mathbf{P}(\mathbf{r})$ [18]. Since there are no magnetic free charges, the bulk and surface magnetic charges are only dipolar in nature, their self-interactions are always present, and there are always equal amounts of positive and negative magnetic charges.